\documentstyle[preprint,aps,eqsecnum,epsf,floats]{revtex}
\tighten

\begin{document}

\newcommand{\dencas}{\mbox{$m_\Lambda - m_\Xi$}}
\newcommand{\denn}{\mbox{$m_\Lambda - m_N$}}
\newcommand{\hf}{\mbox{$h_F$}}
\newcommand{\hd}{\mbox{$h_D$}}
\newcommand{\hc}{\mbox{$h_C$}}
\newcommand{\jay}[2]{\mbox{${\cal J}(\Delta m^{#1}_{#2})$}}
\newcommand{\h}{\mbox{${\cal H}$}}
\newcommand{\kay}[3]{\mbox{${\cal K}(#1,\Delta m^{#2}_{#3})$}}
\newcommand{\gee}[3]{\mbox{$G_{#1,#2,#3}$}}
\newcommand{\geetwid}[3]{\mbox{$\tilde G_{#1,#2,#3}$}}
\newcommand{\be}{\begin{eqnarray}}
\newcommand{\ee}{\end{eqnarray}}
\newcommand{\front}{\mbox{$
{m_K^2  \over 16\pi^2 f_K^2}\ln\left({m_K^2 \over\Lambda_\chi^2}\right)$}}
\newcommand{\llk}{\mbox{$\Lambda\Lambda K^0$}}
\newcommand{\hm}{\mbox{$\xi^\dagger h \xi$}}
\newcommand{\om}{\mbox{$\Omega^-$}}

\def\cV{{\rm V}}
\def\cC{{\cal C}}
\def\cA{{\cal A}}
\def\cW{{\rm W}}
\def\si{{S_{int}}}
\def\cM{{\cal M}}

\def\DKS{[\frac{\partial}{\partial a} K_S]}
\def\DIm{[\frac{\partial}{\partial a} I_3]}

\def\scA{\cA_\si}
\def\scB{\cB_\si}
\def\dm{\Delta m}

\preprint{\vbox{
\hbox{ DUKE-TH-98-172}
}}
\bigskip
\bigskip

\title{Weak Nonleptonic $\Omega^-$ Decay in Chiral Perturbation Theory}
\author{David A. Egolf}
\address{Center for Nonlinear Studies, Theoretical Division (MS-B258)
 Los Alamos National Laboratory, NM 87545.
     \\  {\tt dae@phy.duke.edu} }
\author{Ilarion V. Melnikov}
\address{Laboratory of Atomic and Solid State Physics,
Cornell University, Ithaca, NY 14853.
\\ {\tt lmel@phy.duke.edu}}
\author{Roxanne P. Springer\footnote{Currently on leave at the
    Nuclear Theory Group, University of Washington, Seattle, WA 98195}}
\address{Department of Physics, Duke University, Durham, NC 27708
\\ {\tt rps@phy.duke.edu}}
\maketitle
\begin{abstract}

We calculate the weak nonleptonic decay of the \om \ baryon
to octet final states in heavy baryon chiral perturbation theory (HBChPT).
We include the one-loop leading logarithmic effects and show that this
improves the universality of the HBChPT constants. 
\end{abstract}

\section{Introduction}

The \om \ baryon is the only member of the decuplet of baryons that
decays predominantly through the weak interaction. Its decay to
octet baryons is well measured \cite{PDB} and this provides an
opportunity for finding useful constants in an effective theory
of low energy QCD. We have demonstrated the importance of such
constants for predicting couplings relevant to hypernuclear
decay\cite{llk}.

We will use heavy baryon chiral perturbation theory (HBChPT), described
below, to calculate the \om \ weak decay rate to leading logarithmic
one-loop order.  The constants in HBChPT have been determined to
leading logarithmic one-loop order through
fits to data on axial currents \cite{mj}, the S-wave nonleptonic
decays of the octet of baryons \cite{BSW85a,ej}, and the strong decays
of the decuplet of baryons \cite{bss} to be \cite{fit}
\be\label{wom}
|\cC| = 1.08 \pm 0.05 \, \, , \ \ \ 
D = 0.59 \pm 0.03 \, \,  , \ \ \
F = 0.35 \pm 0.03 \, \, , \ \ \
{\cal H} =-1.76 \pm 0.59 \, \, , \nonumber \\
h_C = 2.38 \pm 9.59 \, \, , \ \ \
h_D = -0.39 \pm 0.22 \, \, , \ \ \ {\rm and} \ \ \ 
h_F = 0.93 \pm 0.47 \, \ \ \ \ .
\ee

Clearly the observables
calculated to date do not constrain  these constants
well enough to provide a meaningful test of the method on any
process sensitive to the large errors in the constants ${\cal H}$,
  $h_C$, $h_D$, and $h_F$.  The 
P-wave nonleptonic decays of the octet baryons and the
$\Lambda \Lambda K$ couplings are both examples of such observables.

One test of how well HBChPT works in describing hadronic observables
is to see how universal the couplings of the theory are.
How well does
nature respect the $SU(3)_L\otimes SU(3)_R$
symmetry?
How well does the chiral limit do as a starting point upon which we
can systematically improve with perturbative loop calculations?
We turn
to the \om \ system for another
test on our description of low energy QCD.

The two body decay modes of the \om\ to the octet are
\begin{eqnarray}
\Gamma(\om \rightarrow \Lambda K^-)&=&(5.42 \pm 0.06) \times 10^{-12}
{\rm MeV}\\
\Gamma(\om \rightarrow \Xi^0 \pi^-)&=&(1.89 \pm 0.06)\times 10^{-12}
{\rm MeV} \\
\Gamma(\om \rightarrow \Xi^- \pi^0)&=&(6.9 \pm 0.3)\times 10^{-13}
{\rm MeV}
\end{eqnarray}
Only two of these modes are independent after isospin symmetry has
been applied, and we will choose the two whose lifetime measurement
for the \om \ are in agreement:
 $\Gamma(\om \rightarrow \Lambda K^-)$
and $\Gamma(\om \rightarrow \Xi^- \pi^0)$.
The tree-level calculation is already in the literature
\cite{ej}.

\section{Heavy Baryon Chiral Perturbation Theory}

Heavy baryon chiral perturbation theory \cite{MJ} is an
effective theory that embodies the symmetries of QCD.
At the low 
momentum transfers involved in the weak decay of the \om,
the mesons and baryons are the relevant degrees of freedom.
Physics which appears in the more fundamental Lagrange density
(QCD)
is mimicked in the effective Lagrange density by a set of
operators and their associated constants.  Were nonperturbative
QCD soluble, we could find these constants by matching the
effective theory onto the full theory. Since this is not
currently possible, we instead use experimental input to
fix the constants.  The power of the effective method is
that it embodies the symmetries of the underlying theory,
provides a power counting which allows calculations to be
performed consistently order by order, and allows \`a priori
estimates of the correction to each order.
It is systematic and model-independent. In this section we
show the parts of the Lagrange density we will need in the
\om decay calculation.  The notation follows that outlined
in \cite{MJ,ej}, except that we choose the convention where the
meson decay constant $f=132$ MeV instead of $f=93$ MeV.
This is compensated by factors of
$\sqrt{2}$ in the Clebsh-Gordan coefficients.

The lowest dimension operators of the
$SU(3)_L\otimes SU(3)_R$ symmetric Lagrange density
will dominate observables,
with the higher dimension operators suppressed by increasing powers
of ${p \over \Lambda_\chi}$, where $p$ is the typical momentum
transfer in the problem and $\Lambda_\chi \sim 1$ GeV is
the chiral symmetry breaking scale.
We will need both weak and strong parts of the Lagrange density
for our one-loop calculation.

\begin{eqnarray}
{\cal L} & = & {\cal L}_{strong} + {\cal L}_{weak}
\ \ \ \  .
\end{eqnarray}
The strong interactions are described by
\begin{eqnarray}\label{strongl}
{\cal L}_{strong} &=& i {\rm Tr} \bar B_v \left(v\cdot {\cal D} \right)B_v
+ 2 D\  {\rm Tr} \bar B_v S_v^\mu \{ A_\mu, B_v \}
+ 2 F\ {\rm Tr}  \bar B_v S_v^\mu [A_\mu, B_v]
\nonumber \\
&&- i \bar T_v^{\mu} (v \cdot {\cal D}) \  T_{v \mu}
+ \Delta m \bar T_v^{\mu} T_{v \mu}
+ {\cal C} \left(\bar T_v^{\mu} A_{\mu} B_v +
  \bar B_v A_{\mu} T_v^{\mu}\right)
\nonumber\\
&& + 2 {\cal H}\  \bar T_v^{\mu} S_{v \nu} A^{\nu}  T_{v \mu}
+ {f^2 \over 8} {\rm Tr} \left( \partial_\mu \Sigma
  \partial^\mu \Sigma^\dagger \right)
+\ \cdots \ \ \ \  ,
\end{eqnarray}
while the $\Delta s=1$ weak interactions
are contained in
\begin{eqnarray}\label{weakl}
{\cal L}_{weak} &=&
G_Fm_\pi^2 f_\pi \Big( h_D {\rm Tr} {\overline B}_v
\lbrace \xi^\dagger h\xi \, , B_v \rbrace \;
+ \;  h_F {\rm Tr} {\overline B}_v
{[\xi^\dagger h\xi \, , B_v ]} \; \nonumber \\ &&
+   h_C {\overline T}^\mu_v
(\xi^\dagger h\xi) T_{v \mu}  \; + \; 
h_\pi {f^2 \over 8} {\rm Tr} \left(  h \, \partial_\mu
\Sigma \partial^\mu \Sigma^\dagger  \right)
 + \ \cdots\ \ \ \Big) \ \ \ \   ,
\end{eqnarray}

\noindent where the lowest mass octet baryons are
\begin{eqnarray}\label{octet}
B_v =
\pmatrix{ {1\over\sqrt2}\Sigma_v^0 + {1\over\sqrt6}\Lambda_v &
\Sigma_v^+ & p_v\cr \Sigma_v^-& -{1\over\sqrt2}\Sigma_v^0 +
{1\over\sqrt6}\Lambda_v&n_v\cr \Xi_v^- &\Xi_v^0 &-
{2\over\sqrt6}\Lambda_v\cr }
\ \ \ ,
\end{eqnarray}

\noindent and the decuplet of baryons are
\begin{eqnarray}\label{decuplet}
& & T^{111}_v  = \Delta^{++}_v, \ \ T^{112}_v =
   {1\over\sqrt{3}}\Delta^{+}_v, 
\ \ T^{122}_v = {1\over\sqrt{3}}\Delta^{0}_v,
  \ \ T^{222}_v = \Delta^{-}_v,
  \nonumber\\
& & T^{113}_v = {1\over \sqrt{3}}\Sigma^{*+}_v,\ \
T^{123}_v  = {1\over\sqrt{6}}\Sigma^{*0}_v, \ \
T^{223}_v = {1\over\sqrt{3}}\Sigma^{*-}_v,
\ \ T^{133}_v = {1\over\sqrt{3}}\Xi^{*0}_v,
\nonumber\\
& & T^{233}_v = {1\over\sqrt{3}}\Xi^{*-}_v, \ \ T^{333}_v =
\Omega^-_v  \ \ \ .
\end{eqnarray}
The subscript $v$ labels the  four-velocity
of the baryon.  The average mass of
the octet of baryons has been explicitly removed from
the Lagrangian \cite{MJ}.  The octet of mesons is contained in
\begin{eqnarray}
\Sigma  = \xi^2= {\rm exp}\left( {2 i M\over f} \right) \ \ \ ,
\end{eqnarray}
where
\begin{eqnarray}
M =
\left(\matrix{{1\over\sqrt{6}}\eta+{1\over\sqrt{2}}\pi^0&\pi^+&K^+\cr
\pi^-&{1\over\sqrt{6}}\eta-{1\over\sqrt{2}}\pi^0&K^0\cr
K^-&\overline{K}^0&-{2\over\sqrt{6}}\eta\cr}\right)
\ \ \ \  .
\end{eqnarray}
The vector and axial vector current
are defined through:
\begin{eqnarray}
V_\mu&=&{1 \over 2} (\xi\partial_\mu\xi^\dagger +
\xi^\dagger\partial_\mu\xi)
\nonumber \\
A_\mu&=&{i \over 2} (\xi\partial_\mu\xi^\dagger -
\xi^\dagger\partial_\mu\xi)
\ \ \ .
\end{eqnarray}

\noindent The covariant chiral derivative is
\be
{\cal D_\mu}= \partial_\mu+[V_\mu,  \; ] \ \ ,
\ee
$\Delta m$ is the average decuplet-octet mass splitting,
and

\begin{eqnarray}
h = \left(\matrix{0&0&0\cr 0&0&1\cr 0&0&0}\right)  \ \ 
\end{eqnarray}
extracts the desired $\Delta s=1$ transition.
The constants appearing above are the axial couplings
$F$, $D$, $\cC$, and ${\cal H}$
which appear in ${\cal L}_{strong}$,
and the weak couplings $h_D$, $h_F$, $h_\pi$, and $h_C$ from
${\cal L}_{weak}$.
The pion decay constant is known to be $f_\pi \sim 132$ MeV.
We have inserted factors of $ G_Fm_\pi^2 f_\pi $ in Eq.~\ref{weakl}\
so that the constants $h_D, h_F$, and $h_C$ are
dimensionless.  The weak interaction Lagrange density is written by
assuming the $\Delta I=1/2$ rule.

In performing our calculations, both here on the \om \ decay and
in previous work on other observables in HBChPT, we retain
the lowest order operators in the
Lagrange density and calculate their
loop corrections.  We keep only the leading logarithmic piece as
this is formally dominant over the operators occurring at the next
order in the Lagrange density (the counterterms).
Numerically, the latter can be
large and competitive with the leading logarithmic result.  To
take into account the error associated with neglecting them, we
include ``theoretical'' error bars at the 25\% level in
performing our fits.  We choose
to do this because the number of constants introduced into the
theory by including the counterterms is prohibitive and  there are
often not enough observables to fix them via experiments.  Some authors
choose to estimate the size of these counterterms using
methods outside of HBChPT, but we prefer to retain a model-independent
result and live with the less constrained predictions we obtain, since
this provides a cleaner test of HBChPT.

\section {Amplitudes and Topologies}

The \om \ decay amplitude to an octet baryon can be written as
\begin{eqnarray}
  \cM(T_v \rightarrow B_v M) =  \overline{u}_B\left(
     \cA^{(P)} \ k^\mu \ +  \  \cA^{(D)}\ k \cdot S_v \ k^\mu\right)
   u_{T \mu} \, \, ,
  \end{eqnarray}
where $T_v$ is a member of the decuplet (in this case the \om)
$B_v$ is an octet final state, and $M$ is a meson.  The $u_B$ and $u_T$ are
baryon spinors, and $S_v$ is the baryon spin operator.  As usual, these
are labelled by velocity quantum numbers.  $\cA^{(P)}$ is
the the P-wave amplitude, and $\cA^{(D)}$ is the D-wave amplitude.
 We will only present the
P-wave calculation in this work since this partial wave dominates
the decay width.
We can also calculate decay parameters but because the experimental
errors are so large they are not useful for testing HBChPT.

\begin{figure}
\epsfxsize=10cm
\hfil\epsfbox{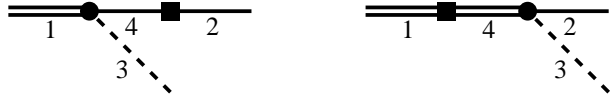}\hfill
\caption{Tree level diagrams contributing to 
  weak \om \ P-wave decay
  to octet baryon final states.  Double lines are decuplet
  baryons, single solid lines are octet baryons, and dashed lines
  are mesons.
The black dots are strong vertices and black squares are 
weak vertices.}
\label{tree}
\end{figure}

The tree level diagrams are shown in Fig.~ \ref{tree}.
The expressions for the tree level amplitudes are \cite{ej}
\be
\cA^{tree}(\om \rightarrow \Lambda K^0) &=& {\cC \over \sqrt{6} f_\pi}\left(
{h_C \over (m_\Omega-m_\Xi^*)}- {h_D - 3h_F \over m_\Xi-m_\Lambda}\right)
\nonumber \\
\cA^{tree}(\om \rightarrow \Xi^0 \pi^-) &=& {h_C \over 3f_\pi}
{\cC \over m_\Omega - m_\Xi^*} \, \, ,
\ee
where we have dropped the P designation since what follows will 
be only P-wave amplitudes.

  We will write the loop level amplitudes as
\begin{eqnarray}
  \cA^{loop}(T_v \rightarrow B_v M) = C^N \left[\cV^N + { 1 \over 2}
    \cW^N \right] \, \, ,
  \end{eqnarray}
  where $N$ designates the diagram number from Figs.~\ref{loop1} and
\ref{loop2},
  and we have separated the contributions to the vertex renormalization,
  $\cV$, from those to the
 wavefunction renormalization,
$\cW$.
  We then calculate the width using
\begin{eqnarray}
\Gamma(T \rightarrow B \pi) &=&
{|\vec{k}| \over 6 \pi} {m_{B} \over  m_{T}}
\left[{\left(
     m_{T} - m_{B}\right)}^2 - m^2_{M}\right]\nonumber\\
&&\times \ {\left|  \cA^{tree} + \sum_{N} C^N
         \left(\cV^N + {1 \over 2} \cW^N \right)
         \right|}^2,
\end{eqnarray}
where $m_T$, $m_B$, and $m_M$ are the masses of the decuplet,
octet, and meson,
respectively, $\vec{k}$ is the 3-momentum of an outgoing particle, and the
sum is over all diagrams $N$.  The Clebsh-Gordan coefficients for
each graph are contained in $C^N$, where we also include factors of
the meson decay constant $f_\pi$.

Below we give the expressions for nonzero $\cV^N$ and
$\cW^N$ contributions
in terms of the labels in Figs.~ \ref{loop1} and \ref{loop2}, and
the following definitions

\begin{eqnarray}
I(m)    &=&  - \  {i \over 16 \pi^2}
              m^2 \ln \left({m^2 \over \Lambda^2_\chi}\right) \\
\tilde{I}(m_1, m_2) &=&  {m_1^2 \over m_1^2-m_2^2} I(m_1)+
                         {m_2^2 \over m_2^2-m_1^2} I(m_2)\\
Q(a, m)   &=&  {i \over 48 \pi^2}
               \left[
                     \left(2a^2 - 3m^2\right)~a
                     \ln \left({m^2 \over \Lambda^2_\chi}\right) \right.
                   \nonumber\\
          &&+        \left. 2 {(a^2 - m^2)}^{3 \over 2}
                     \ln  \left({a + \sqrt{a^2 -m^2 + i \bar{\epsilon}}\over
                                 a - \sqrt{a^2 -m^2 + i \bar{\epsilon}}
                                }
                         \right)
                     -2 \pi m^3
               \right]\\
\tilde{Q}(a, m_1, m_2)  &=&   {m_1^2 \over m_1^2-m_2^2} Q(a, m_1)+
                              {m_2^2 \over m_2^2-m_1^2} Q(a, m_2)\\
Q'(a,m)  &=&  {\partial \over \partial a} Q(a,m)\nonumber\\
         &=&  {i \over 16 \pi^2}
              \left[
                     \left(2a^2 - m^2\right)
                     \ln \left({m^2 \over \Lambda^2_\chi}\right) \right.
\nonumber\\
         &&+        \left. 2 a \sqrt{a^2 - m^2}
                     \ln  \left({a + \sqrt{a^2 -m^2 + i \bar{\epsilon}}\over
                                 a - \sqrt{a^2 -m^2 + i \bar{\epsilon}}
                                }
                          \right)
             \right]\\
\tilde{Q}'(a, m_1, m_2)   &=&  {m_1^2 \over m_1^2-m_2^2} Q'(a, m_1)+
                               {m_2^2 \over m_2^2-m_1^2} Q'(a, m_2)
\end{eqnarray}

The functional mass dependence is given explicitly.
The subscripts on the
masses correspond to the labels on the particle lines shown in
Figs.~\ref{loop1} and \ref{loop2}.  We include the full
nonzero decuplet-octet mass splitting dependence ($\Delta m$) in
our formulas.  The contributions to the vertex renormalization are:

\begin{eqnarray}
\cV^{( 2)}  \, = \, - \ { I(m_5) \over m_2 - m_4} \ \ \ \ \ \ &&
 \cV^{( 4)}  \, = \, - \ { I(m_5) \over m_2 - m_4} \nonumber \\
 \cV^{( 6)}  \, = \,  \tilde{I}( m_4, m_5)     \ \ \ \ \ \ &&
 \cV^{( 7)}  \, = \, {I(m_5) \over m_1 - m_4}   \nonumber \\
 \cV^{( 9)}  \, = \,  {I(m_5) \over m_1 - m_4}  \ \ \ \ \  \ && 
 \cV^{( 64)} \, = \, - \ {1 \over 2 \, \dm } \
 {Q(\dm, m_7) \over m_2 - m_6}
                            \nonumber \\
 \cV^{( 66)} \, = \,   {\tilde{Q}(\dm, m_7, m_6) \over 2 \dm} \ \ \ \ \ \ &&
 \cV^{( 68)} \, = \, - \ {3 \over 4} \ {I(m_6) \over m_2 - m_4}
 \nonumber \\
 \cV^{( 72)} \, = \, {\tilde{Q}(\dm, m_5, m_7) \over m_1 - m_6}
 \ \ \ \ \ \ &&
 \cV^{( 73)} \, = \,  {\tilde{Q}(\dm, m_5, m_7) \over 3 \, \dm}
 \nonumber \\
 \cV^{( 75)} \, = \, {2 \, \tilde{Q}(\dm, m_6, m_7) \over m_2 - m_4}
       \ \ \ \ \ \ &&
 \cV^{( 77)} \, = \, - \  {1 \over m_2 - m_7} \ 
                    {Q(\dm, m_5) \over 3 \, \dm}       \nonumber \\
 \cV^{( 82)} \, = \,  {2 \, Q'(\dm, m_6) \over m_2 - m_4} \
 \ \ \ \ \ \ &&
 \cV^{( 86)} \, = \, - \ {1 \over m_1 - m_6} \ Q'(\dm, m_5)
 \nonumber \\
 \cV^{( 90)} \, = \,  {5 \over 12} \ {I(m_6) \over m_2 - m_5}
 \ \ \ \ \ \ &&
 \cV^{( 94)} \, = \, - \ {5 \over 12} \ \tilde{I}( m_7, m_5)
 \nonumber \\
 \cV^{( 102)} \, = \, - \ {5 \over 18} \
 {\tilde{Q}(\dm, m_7, m_5) \over \dm} \ \ \ \ \ \ &&
 \cV^{( 104)} \, = \, {5 \over 18} \ {1 \over m_2 - m_5}
                   {Q(\dm, m_6) \over \dm}          \nonumber \\
 \cV^{( 118)} \, = \,   {5 \over 12} \ {I(m_6) \over m_1 - m_4}
 \ \ \ \ \ \ &&
 \cV^{( 121)} \, = \, {1 \over m_1 - m_7} \ 
                   {Q(\dm, m_6) \over 2 \,\dm}         \nonumber \\
 \cV^{( 124)} \, = \,  {1 \over m_1 - m_7} \ 
                   {Q(\dm, m_6) \over 3 \, \dm}          \ \ \ \ \ \ &&
 \cV^{( 126)} \, = \, - \ {5 \over 12} \ {I(m_5) \over m_1 - m_7}
 \nonumber \\
 \cV^{( 129)} \, = \, - \ {5 \over 18} \ {1 \over m_1 - m_7} \
                     {Q(\dm, m_5) \over \dm}          \ \ .
\end{eqnarray}

\noindent Two of the contributions require more care:
\begin{eqnarray}
 \cV^{(60)} &=&  - {3 \over 4} \tilde{I}  (m_6, m_7)
 \mbox{~~ if~~} m_4-m_2  \approx m_s           \nonumber\\
 \cV^{103}&=& -{5 \over 12} \tilde{I} (m_6, m_7)
 \mbox{~~ if~~} m_1 - m_5 \approx m_s          \nonumber
\end{eqnarray}
\noindent and otherwise contribute zero at this order.

The contributions to the wavefunction renormalization are
\be
 \cW^{( 30)} \, = \, {i \, I(m_6) \over m_2 - m_4} \ \ \ \ \ \ &&
 \cW^{( 52)} \, = \, - \ {i \, I(m_6)\over m_1 - m_4}     \nonumber \\
 \cW^{( 60)} \, = \, - \ {3 \over 4} \ \tilde{I}(m_6, m_7) \ \ \ \ \ \ &&
 \cW^{( 62)} \, = \, {3 \over 4} \ {I(m_6) \over m_2 - m_7} \nonumber \\
 \cW^{( 72)} \, = \, - \  \tilde{I}(m_5, m_7)             \ \ \ \ \ \ &&
 \cW^{( 75)} \, = \, - \ 2 \ \tilde{Q}'(\dm, m_6, m_7)      \nonumber \\
 \cW^{( 76)} \, = \, {I(m_5) \over m_2 - m_7}           \ \ \ \ \ \ &&   
 \cW^{( 78)} \, = \, - \ 2 \ {Q'(\dm, m_6) \over m_2 - m_7}
 \nonumber \\
 \cW^{( 80)} \, = \, - \ {I(m_5) \over m_1 - m_7}            \ \ \ \ \ \ &&
 \cW^{( 83)} \, = \, {3 \over 4} \  {I(m_6) \over m_2 - m_7}   \nonumber \\
 \cW^{( 88)} \, = \, 2 \ {Q'(\dm, m_7) \over m_2 - m_4}      \ \ \ \ \ \ &&
 \cW^{( 103)} \, = \, - \ {5 \over 12} \ {\tilde{Q}'(\dm, m_6, m_7)}
     \nonumber \\
     \cW^{( 106)} \, = \, {5 \over 12} \ {Q'(\dm, m_7) \over m_2 - m_5}
         \ \ \ \ \ \ &&
 \cW^{( 112)} \, = \, -  \ {5 \over 12} \ {Q'(\dm, m_6) \over m_1 - m_4}
     \nonumber \\
 \cW^{( 119)} \, = \, - \ {3 \over 4} \ {I(m_6) \over m_1 - m_7}
     \ \ \ \ \ \ &&
 \cW^{( 122)} \, = \, - \ 2 \ {Q'(\dm, m_6) \over m_1 - m_7}     \nonumber \\
 \cW^{( 123)} \, = \, - \ {I(m_6) \over m_1 - m_7}       \ \ \ \ \ \ &&
 \cW^{( 130)} \, = \, - \ {5 \over 12} \ {Q'(\dm, m_6) \over m_1 - m_7}
 \ \ \ .
\ee

We have shown above the integrals generated by each of the
diagrams in Fig. \ref{loop1} and \ref{loop2} but not the
associated Clebsh-Gordan coefficients.  Because there are
about 500 intermediate states, it is not feasible to
present them here.  

\section{Results and Conclusions}

Our expressions for the \om \ particle decay, along with the
experimental width measurements for each decay mode,
were included in a global fit to
the HBChPT theory constants $\cC$, $D$, $F$, ${\cal H}$, $h_C$
$h_D$, and $h_F$.  We obtained
\be
|\cC| = 1.08 \pm 0.05 \, \, , \ \ \
D = 0.58 \pm 0.03 \, \, , \ \ \ 
F = 0.35 \pm 0.03 \, \, , \ \ \
{\cal H} =-1.90 \pm 0.50 \, \, , \\ \nonumber
h_C = 2.93 \pm .43 \, \, , \ \ \
h_D = -0.50 \pm 0.16 \, \, , \ \ \ {\rm and} \ \ \ 
h_F = 0.91 \pm 0.10\, \, . \ \ \
\ee

Though our expressions are valid for finite decuplet-octet splitting
the fit was done taking $\Delta m=0$.
This is necessary because many of the other calculations involved in the
fit were performed in the $\Delta m \rightarrow 0$ limit. It would be just
as inconsistent to attempt to compare calculations done with
$\Delta m=0$ to those done with $\Delta m \ne 0$ in a global
fit as it would to compare calculations done at tree level
to those done to loop level. One of the reasons  this consistency
is so important is that SU(3) symmetric effects can generate
large differences in the HBChPT coefficients fit at different
orders even though differences in actual observables remain stable.

Note the dramatic improvement in the determination of the constant
$h_C$ through the inclusion of the \om \ decay rates.
The constants of HBChPT 
 are not physically meaningful by themselves.  However, they are
useful in two ways.  First, the goodness of fit to the
available leading logarithmic one-loop calculations and the decrease
in error bars from the constants shown in the fit without including
the \om \ (Egn.~\ref{wom}) show that
these systems are well described by a perturbative
calculation in HBChPT.
Second, these constants are necessary for
making predictions and performing further tests of HBChPT.  We can
still expect some calculations of observables in HBChPT
to show an unfortunate propagation of errors that
 will make HBChPT not restrictive for those observables, but the
 tests will be more meaningful now that the errors on the constants have
 been dramatically reduced.

 We are now in a position to make predictions on other
 observables involved in weak \om \ decay.  Its weak decay to the decuplet
 of baryons has been measured, and there is some data on asymmetry
 parameters, which makes it useful to calculate the other partial
 waves.  There is some evidence that the $\Delta I=1/2$ rule
 is violated by the \om \ decays \cite{georgi}.  We do not test that
 here beyond having assumed the $\Delta I=1/2$ rule and getting
 a good fit to the two isospin independent decay modes.  We think
 our results are evidence in favor of the utility of HBChPT applied
 to hadronic observables.

\bigskip
\bigskip

\centerline{Acknowledgements}

\bigskip

The authors thank the Institute for Nuclear Theory at the University
of Washington for their hospitality while some of this work was
completed.  DAE and IVM thank the Universit\"at  Bayreuth Physikalisches
Institut for
hosting their visit during this work.
RPS acknowledges support from the Department of Energy
under grant no.  DE-FG02-96ER40945.   DAE and IVM acknowledge support by
the
National Science Foundation grant no. DMR-9705410.  RPS thanks Martin
Savage for useful discussions.

\begin{figure}
\epsfxsize=15cm
\hfil\epsfbox{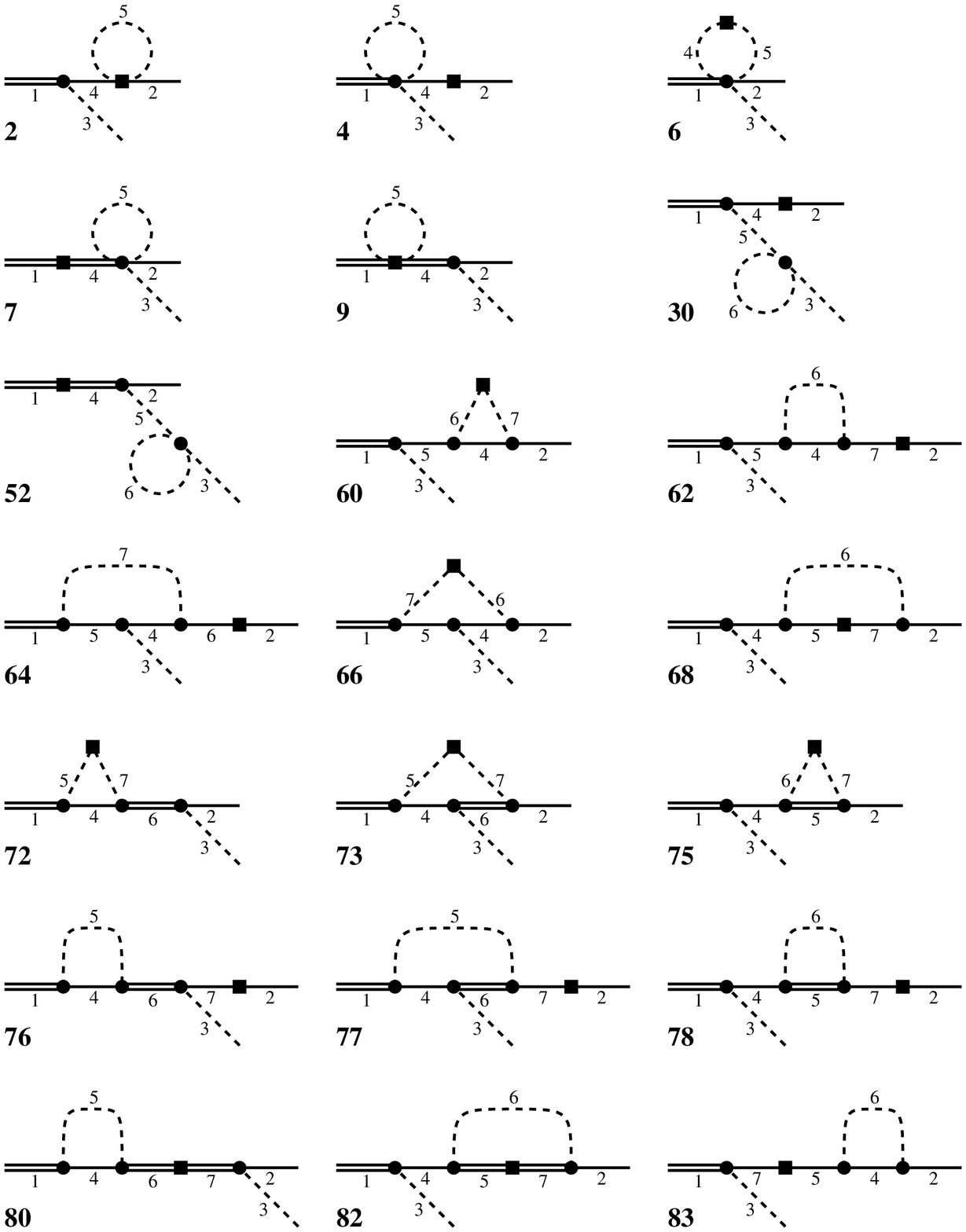}\hfill
\caption{Loop level diagrams contributing to 
  weak \om \ P-wave decay to octet baryon final states.
   Double lines are decuplet
  baryons, single solid lines are octet baryons, and dashed lines
  are mesons.
The black dots are strong vertices and black squares are 
weak vertices.}
\label{loop1}
\end{figure}

\begin{figure}
\epsfxsize=15cm
\hfil\epsfbox{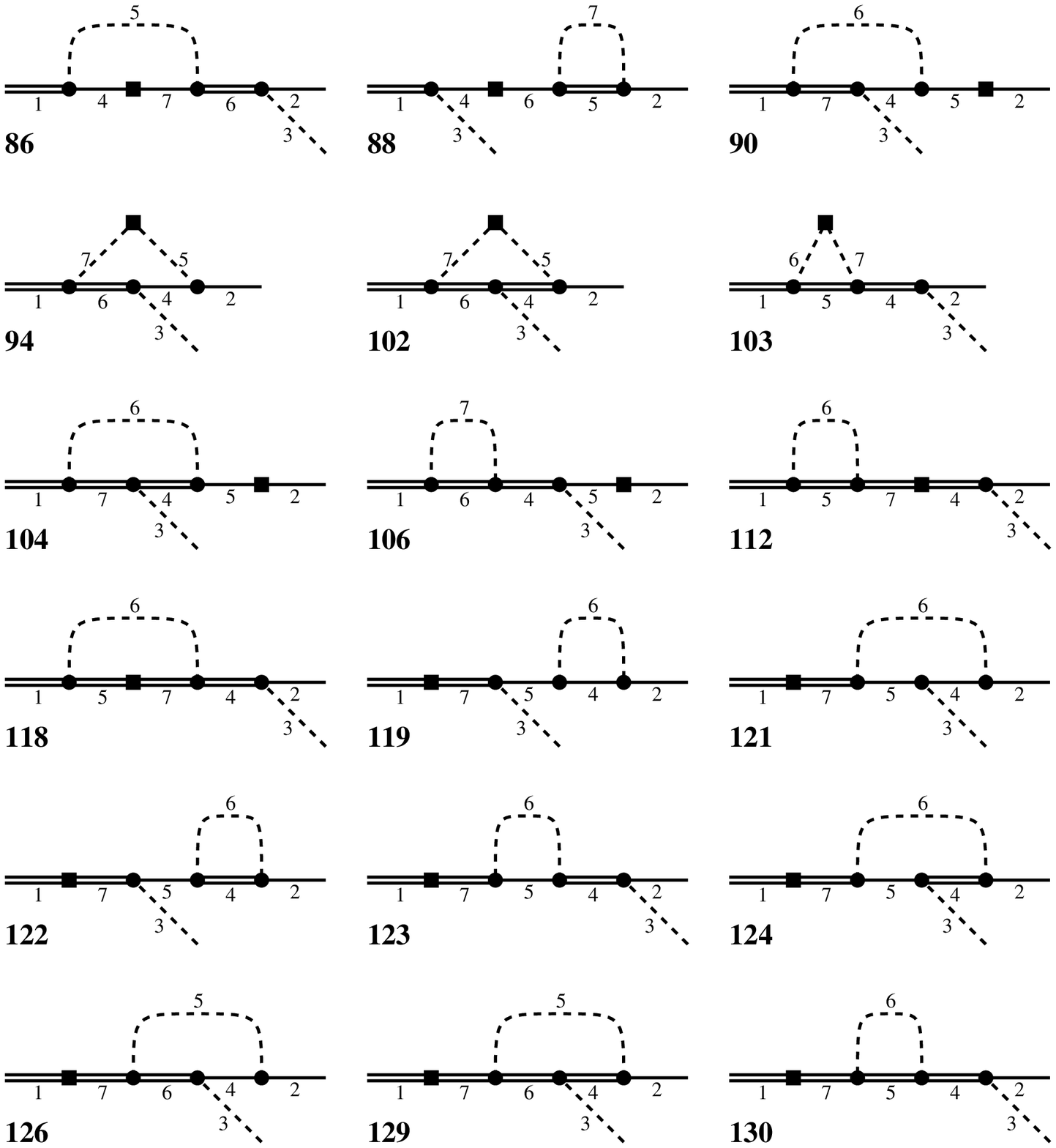}\hfill
\caption{Continuation of loop level diagrams contributing to 
  weak \om \ P-wave decay to octet baryon final states.
   Double lines are decuplet
  baryons, single solid lines are octet baryons, and dashed lines
  are mesons.
The black dots are strong vertices and black squares are 
weak vertices.}
\label{loop2}
\end{figure}


\begin{references}

\bibitem{PDB} Particle Data Book, Phys. Rev. {\bf D54}, 1 (1996).

\bibitem{llk} M.J. Savage and R.P. Springer, Phys. Rev. {\bf C57},
  1478 (1998)

  \bibitem{MJ} The formalism of heavy baryon chiral perturbation
theory is introduced in E. Jenkins and A.V. Manohar,
In ``Dobogokoe 1991, Proceedings, 
{\sl Effective field theories of the standard model}'' 113-137.

\bibitem{mj} E. Jenkins and A. Manohar, Phys. Lett. {\bf B255}, 558 (1991); 
 {\bf B259}, 353 (1991).


\bibitem{BSW85a} J. Bijnens, H. Sonoda and M.B. Wise, 
Nucl. Phys. {\bf B261}, 185 (1985).

\bibitem{ej} E. Jenkins, Nucl. Phys. {\bf B375} 561 (1992).

  
\bibitem{bss} M.N. Butler, M.J. Savage and  R.P. Springer, Nucl. 
Phys. {\bf B399}, 69 (1993).


\bibitem{fit} M.J. Savage and  J. Walden, 
Phys. Rev. D {\bf 55} 5367 (1997); R.P. Springer, DUKE-TH-97-152.

\bibitem{georgi} C. Carone and H. Georgi, Nucl. Phys. {\bf B375} 243 (1992).
\end{references}
\end{document}